\RequirePackage{fix-cm}
\documentclass[twocolumn]{svjour3}          
\smartqed  
\usepackage{graphicx}
\usepackage[hidelinks]{hyperref}
\usepackage[utf8]{inputenc}
\usepackage[T1]{fontenc}
\usepackage{csquotes}
\newcommand*{\affaddr}[1]{#1} 
\newcommand*{\affmark}[1][*]{\textsuperscript{#1}}
\newcommand*{\Letter}{\relax}
\begin{document}

\title{Software Training in HEP
}


\author{
Sudhir Malik  (editor)~\protect\affmark[1] 
\and
Samuel Meehan \affmark[2]
\and
Kilian Lieret\affmark[3] 
\and
Meirin Oan~Evans\affmark[4]
\and
Michel H.\ Villanueva\affmark[5]
\and
Daniel S.\ Katz\affmark[6] 
\and
Graeme A.\ Stewart\affmark[2] 
\and
Peter Elmer\affmark[7]
\and
Sizar Aziz\affmark[8]
\and
Matthew Bellis\affmark[18] 
\and
Riccardo Maria Bianchi\affmark[25] 
\and
Gianluca Bianco\affmark[30,31]
\and
Johan Sebastian Bonilla\affmark[23]
\and
Angela Burger\affmark[24] 
\and
Jackson Burzynski\affmark[27]
\and
David Chamont\affmark[8] 
\and
Matthew Feickert\affmark[6]
\and
Philipp Gadow\affmark[12] 
\and
Bernhard Manfred Gruber\affmark[2,35,36] 
\and
Daniel Guest\affmark[15]
\and
Stephan Hageboeck\affmark[2] 
\and
Lukas Heinrich\affmark[2]
\and
Maximilian M. Horzela\affmark[16]
\and
Marc Huwiler\affmark[26]
\and
Clemens Lange\affmark[2] 
\and
Konstantin Lehmann\affmark[17] 
\and
Ke Li\affmark[19]
\and
Devdatta Majumder\affmark[28]
\and
Judita ~Mamu{\v{z}}i{\'{c}}\affmark[10] 
\and
Kevin Nelson\affmark[22] 
\and
Robin Newhouse\affmark[13]
\and
Emery Nibigira\affmark[14]
\and
Scarlet Norberg\affmark[1]
\and
Arturo S\'{a}nchez Pineda\affmark[11] 
\and
Mason Proffitt\affmark[19] 
\and
Brendan Regnery\affmark[23]
\and
Amber Roepe\affmark[34] 
\and
Stefan Roiser\affmark[2] 
\and
Henry Schreiner\affmark[7]
\and
Oksana Shadura\affmark[21] 
\and
Giordon Stark\affmark[9] 
\and
Stephen Nicholas Swatman\affmark[2,20]
\and
Savannah Thais\affmark[7]
\and
Andrea Valassi\affmark[2] 
\and
Stefan Wunsch\affmark[2,16]
\and
David Yakobovitch\affmark[32]
\and
Siqi Yuan\affmark[29]
}

\authorrunning{The HSF Training WG}

\institute{
\Letter Sudhir Malik\\
sudhir.malik@upr.edu\\
\\
\affaddr{\affmark[1]University of Puerto Rico Mayaguez, USA}\\
\affaddr{\affmark[2]CERN, Geneva, Switzerland}\\
\affaddr{\affmark[3]Ludwig Maximilian University of Munich, Germany}\\
\affaddr{\affmark[4]University of Sussex, Brighton, UK}\\
\affaddr{\affmark[5]University of Mississippi, Oxford, MS, USA}\\
\affaddr{\affmark[6]University of Illinois at Urbana-Champaign, IL, USA}\\
\affaddr{\affmark[7]Princeton University, Princeton, NJ, USA}\\
\affaddr{\affmark[8]Universit\' e Paris-Saclay, CNRS/IN2P3, IJCLab, Orsay, France}\\
\affaddr{\affmark[9]Santa Cruz Institute for Particle Physics, UC Santa Cruz, CA, USA}\\
\affaddr{\affmark[10]Instituto de F\' ssica Corpuscular / Consejo Superior de Investigaciones Cient\' ificas - University of Valencia (IFIC / CSIC - UV), Spain }\\
\affaddr{\affmark[11]LAPP, Universit\'e Savoie Mont Blanc, CNRS/IN2P3, Annecy, France
}\\
\affaddr{\affmark[12]Deutsches Elektronen-Synchrotron DESY, Hamburg, Germany }\\
\affaddr{\affmark[13]University of British Columbia, Vancouver, Canada }\\
\affaddr{\affmark[14]Universit\'e de Strasbourg, CNRS, IPHC UMR 7178, Strasbourg, France }\\
\affaddr{\affmark[15]Humboldt University of Berlin, Berlin, Germany }\\
\affaddr{\affmark[16]Karlsruhe Institute of Technology (KIT), Karlsruhe, Germany }\\
\affaddr{\affmark[17]Simon Fraser University, Burnaby BC, Canada }\\
\affaddr{\affmark[18]Siena College, Loudonville, NY, USA }\\
\affaddr{\affmark[19]University of Washington, Seattle, WA, USA }\\
\affaddr{\affmark[20]University of Amsterdam, Amsterdam, The Netherlands }\\
\affaddr{\affmark[21]University of Nebraska, Lincoln, NE, USA }\\
\affaddr{\affmark[22]University of Michigan, Ann Arbor, MI, USA }\\
\affaddr{\affmark[23]University of California Davis, Davis, CA, USA }\\
\affaddr{\affmark[24]Oklahoma State University, Stillwater, OK 74078, USA }\\
\affaddr{\affmark[25]University of Pittsburgh, Pittsburgh, PA, USA }\\
\affaddr{\affmark[26]Universität Zürich, Zürich, Switzerland }\\
\affaddr{\affmark[27]University of Massachusetts, Amherst MA, USA }\\
\affaddr{\affmark[28]Institut Ruđer Bo\v skovi\' c, Zagreb, Croatia }\\
\affaddr{\affmark[29]Boston University, Boston, USA }\\
\affaddr{\affmark[30]University of Bologna, Bologna, Italy }\\
\affaddr{\affmark[31]INFN (Istituto Nazionale di Fisica Nucleare), Italy }\\
\affaddr{\affmark[32]SingleStore, New York}\\
\affaddr{\affmark[34]University of Oklahoma, Norman, OK, USA } \\
\affaddr{\affmark[35]Center for Advanced Systems Understanding, Saxony, Germany}\\
\affaddr{\affmark[36]Technische Universität Dresden, Dresden, Germany }\\
}

%
%


\date{Received: 1 August 2021 / Accepted: date}
\maketitle
\begin{abstract}
The long term sustainability of the high energy physics (HEP) research software ecosystem is essential to the field. With new facilities and upgrades coming online throughout the 2020s, this will only become increasingly important. Meeting the sustainability challenge requires a workforce with a combination of HEP domain knowledge and advanced software skills. The required software skills fall into three broad groups. The first is fundamental and generic software engineering (e.g., Unix, version control, C++, continuous integration). The second is knowledge of domain-specific HEP packages and practices (e.g., the ROOT data format and analysis framework). The third is more advanced knowledge involving specialized techniques, including parallel programming, machine learning and data science tools, and techniques to maintain software projects at all scales. This paper discusses the collective software training program in HEP led by the HEP Software Foundation (HSF) and the Institute for Research and Innovation in Software in HEP (IRIS-HEP). The program equips participants with an array of software skills that serve as ingredients for the solution of HEP computing challenges. Beyond serving the community by ensuring that members are able to pursue research goals, the program serves individuals by providing intellectual capital and transferable skills important to careers in the realm of software and computing, inside or outside HEP.
\end{abstract}

\section{Introduction}
\label{intro}
Particle physics in the coming decades will continue to explore the fundamental workings of the universe. This requires  upgrading existing major facilities like the Large Hadron Collider (LHC) to the High Luminosity LHC~\cite{hl-lhc} and building new facilities like the Long-Baseline Neutrino Facility (LBNF)~\cite{papadimitriou2017design} and Deep Underground Neutrino Experiment (DUNE)~\cite{dune}, among many others.  To realise the full physics potential of this work, an equivalent investment must be made into the software required to collect, process, and analyse the deluge of the data recorded. Recent efforts like the HSF~\cite{HSF-homepage} and IRIS-HEP~\cite{iris-hep} are facilitating cooperation and common efforts in HEP software and computing worldwide to develop state-of-the-art software cyberinfrastructure required to meet the challenges of the upcoming HEP experiments'  data-intensive scientific research. The rapid evolution of computing technology with a  concomitant increase in the complexity of software algorithms for analysis requires developers to acquire a broad portfolio of programming skills in order to enable future discoveries.

It is critical that all stakeholders across HEP make a major effort to provide a strong foundation for new researchers entering the field. The researchers must be brought up to date with new software technologies, concurrent programming, and artificial intelligence, and must maintain, improve, and sustain the existing HEP software. However, young researchers graduating from universities worldwide currently do not receive adequate preparation in modern computing practices to respond to the growing needs related to the above experimental challenges. A community white paper \cite{hsfwptraining} outlined the initiatives to address training needs and issues that need to be taken into account for these to be successful. In the last two years, the HSF Training working group, together with IRIS-HEP and FIRST-HEP~\cite{first-hep} and partnering with The~Carpentries~\cite{carpentries}, has begun development of a software training program. The efforts of this group have been focused on two specific goals: (1) developing material for an introductory HEP software curriculum, and (2) teaching this curriculum to HEP scientists. Thus far, over 1000 people in HEP and related computing areas have been trained. This paper describes the activities, the curriculum, and future directions of HEP software training.
\section{Organization }\label{sec:Organisation}
The HSF Training working group, which is led by three co-conveners, engages with different experimental collaborations and initiatives such as IRIS-HEP, FIRST-HEP, and The Carpentries. The training group has weekly public meetings~\cite{HSF-training-meetings} to plan and assess progress. This is where ideas and proposals are discussed and events are planned. These meetings are held remotely using Zoom and live notes are maintained for anyone unable to join. Training events are announced via several email lists, with registration and timetables organized using  Indico~\cite{HSF-training-events}.

The style and pedagogy of the training is heavily inspired by The Carpentries.
The training is student-centric, suitable for self-study, and experiment agnostic, with reusable study material that is open source and open access, and hosted in the HSF's Training repositories on  GitHub~\cite{HSF-training-materials}. We encourage participants to provide feedback and suggestions for improvement by opening issues in these repositories or to directly help with the development by opening pull requests.
In most cases the training material is in the form of a website that is built from files written in the easy-to-learn Markdown language. The website is automatically built using the static site generator Jekyll~\cite{jekyll} via GitHub Pages~\cite{github-pages} and an adapted and extended template from The Carpentries~\cite{hsf-styles,carpentry-styles}. Thus the entry barrier to contribute to the material is fairly low, as only basic knowledge of git is required (and in most cases, all necessary steps can be performed via the GitHub web interface).
All lessons are listed in the HSF Training Center~\cite{HSF-curriculum}, which provides an overview of the available training modules and serves as an entry point for anyone wishing to learn.

Based on our experiences, we have also formalized the procedure used to organize  a training event and have compiled our knowledge in a compact guide~\cite{HSF-training-how-to}.
As organization is all about dividing work, we distinguish between three relevant roles at our events:
\begin{itemize}
    \item \emph{Instructors} are subject-matter experts who develop training material and then teach it, either in person, in recorded live sessions, or by recording videos before the event.  Instructors are the primary academic drivers of the program at large and provide guidance to mentors and students alike. They gain experience in curriculum design with a focus on optimizing pedagogy for all learning styles.
    \item \emph{Mentors} work closely with participants, for example, by conducting small group mentoring sessions with ideally only five students per mentor. They optimize the learning environment for individual participants and help them persevere.  They are critical to the success of any event and through participation as a mentor not only serve the community, but develop pedagogical communication skills that are transferable to other aspects of their research/teaching portfolio.
    \item \emph{Facilitators} take care of organizational aspects. They are responsible for putting together all of the pieces of the puzzle to successfully execute the full event while serving as the primary point of reference for participants to communicate. They take on a dynamic responsibility beyond the \enquote{core content} of the training event itself, and they also learn the essential \enquote{soft skills} necessary to be a leader in the academic community and beyond.
\end{itemize}
All three groups are collectively referred to as \emph{educators}.
As creating training material and teaching requires a lot of commitment and time, it is of great importance to acknowledge the efforts of everyone involved. Currently this is is mostly achieved by listing helping community members on the pages of the relevant training and on a central community page~\cite{HSF-training-community}.

Finally, Blueprint workshops~\cite{HSF-training-blueprint} and hackathons \cite{HSF-training-hackathon} are organised to brainstorm new training events, develop content, and discuss improvements. The travel cost for educators and video captioning of training material have been supported by IRIS-HEP and FIRST-HEP.

\section{Curriculum}\label{sec:Curriculum}

An initial survey of the software and training needs of the HEP community was conducted in February of 2019 \cite{david_lange}. This was followed by the development of \enquote{prototype} course modules and  pilot training events from which feedback from participants was solicited.

Based on the surveys and the experiences gathered at the events, the course structure was extended into a full curriculum consisting of a variety of training modules.
Each training module is independent from the others (except for some clearly marked requirements), so that students can prioritize certain skills before others. This is especially important in academia because students are often expected to work directly towards scientific results with minimal time given for acquiring software knowledge or best practices.

The most basic skill set (Unix shell, Python, git) is covered by modules directly developed by the Software Carpentry~\cite{softcarp}. A large module that covers the basics of modern C++ is currently in development and other modules focusing on development in C++ such as CMake have already been taught with great success.

This is complemented by a series of broader software engineering topics, such as continuous integration and deployment using both GitHub Actions and GitLab CI as examples. These modules are also particularly relevant for analysis preservation, for which modules covering domain specific software such as REANA~\cite{reana} are in development.

A lesson on machine learning and a lesson specifically targeting machine learning with GPUs started a data analysis techniques curriculum section. Similarly important are HEP specific tools, especially ROOT~\cite{root_cern} and integration such as uproot~\cite{uproot}.

Finally, development is ongoing of modules that cover advanced topics that are important for students striving to become core developers, such as code documentation, performance optimization and parallel programming.

The module list~\cite{HSF-curriculum} and the material evolves continuously depending on input from participants and person-power available; as it is open source, any interested stakeholder can contribute.

\section{Training Events}\label{sec:Training}

During the initial period of training, 150 people received \enquote{introductory} software skills training at Fermilab (FNAL), Argonne National Lab (ANL), Lawrence Berkeley Lab (LBNL), and CERN~\cite{HSF-training-SC-FNAL-2019,HSF-training-ATLAS-ANL-2019,HSF-training-ATLAS-LBNL-2019,HSF-training-SC-CERN-2019}. National labs are the hub of the HEP community and provide an environment where it is easier to reach a diverse population of participants with good infrastructure for in-person training. At the CoDaS-HEP school~\cite{codas-hep}, over 50 people participated in the advanced \enquote{computing bootcamp} software training. These training events were in-person.

However, the COVID-19 pandemic necessitated a rapid adjustment to virtual platforms, which evolved throughout the course of 2020 as we gained experience. The events that we had to pivot to use a virtual environment include training on continuous integration and deployment~\cite{cicd_gitlab,cicd_github}, Docker~\cite{docker}, machine learning on GPUs \cite{mlgpu} and C++~\cite{cplusplus1} (organized together with SIDIS~\cite{sidis}).

To date, nearly 100 educators have taught over 1000 participants in about a dozen training events. Valuable lessons have been learned regarding in-person and virtual training. There is very clear and detailed guidance for anyone willing to host, request or organize a training while staying aligned with the approach, philosophy, and code of conduct of the HSF Training group so as to make the tools and techniques that are developed persistent, reuseable, and broadly accessible~\cite{HSF-training-how-to}.

While in-person events offer more opportunities for active and efficient engagement of participants and community building, they are generally more exclusive: Participants need sufficient funding and extra preparation time to arrange travel to the venue. Hosts have to book specially arranged/equipped rooms with multiple projectors and screens to simultaneously show teaching materials and slides. The space constraints typically limit the number of participants to a few dozen and a long lead time is required for the logistics.
Our in-person events have been managed by about five educators, which is necessary for the \enquote{hands-on} aspect to be successful. These educators also need to make a large time commitment; they cannot just present their material and leave.
Virtual events have a broader reach of participant attendance that is much higher compared to in-person events and enable a considerably more equitable service to the community. Because the teaching materials are fully preserved via lesson creation and YouTube videos beforehand, an inability to attend during the scheduled time does not considerably degrade learning.  Finally, these video materials are captioned to be inclusive of those with hearing impairments. Captioning videos for a week-long event ($\sim$\$50/day) is considerably more economical than the cost of a hired sign language interpreter ($\sim$\$1000/day).

The disadvantage of virtual events, however, is that it is difficult for educators and participants to interact closely -- you just can’t recreate the in-person environment on Zoom. Educators and participants have to plan and act upon their spread across time zones in the best possible way. It is also challenging to keep everyone engaged and on the same page due to the pervasive culture of \enquote{multi-tasking} within HEP. Due to this issue, although initial registrations for these events are very high, the actual attendance is typically only 50\% of those who have registered. The online experience is more prone to be distracted by other professional duties.
However, it should be noted that this does not mean that there is a lesser degree of learning occurring at the training event. Tools like Mattermost, discord, and Slack have been effectively deployed for asynchronous communication, both during and after the event.

In general, devoting full time to training is always challenging because of a combination of two factors.  Though there is widespread desire to engage in training, there is an institutional culture that prioritizes immediate research activity over dedicated professional development, even though the latter will lead to higher productivity in the long term.
\section{Feedback}\label{sec:Feedback}
Feedback is required for us to evaluate if we are effectively facilitating learning and to ensure the success of future training. Every training has a pre- and post- survey to collect feedback from the participants. This includes a set of baseline questions pertaining to demographics and questions to assess the quality and method of training. These questions can be adapted to the nature and topic of each training event. Additionally we organize a \enquote{post-mortem} session among the educators to internally discuss the successes and failures. This typically occurs after completion of the results of the post- (and pre-) workshop surveys, which guide the discussion. Finally, a short presentation about the training experience is presented at the HSF Training weekly meeting and/or at the HSF all-working-groups planning meeting.

\begin{figure*}
\centering
\includegraphics[width=11cm]{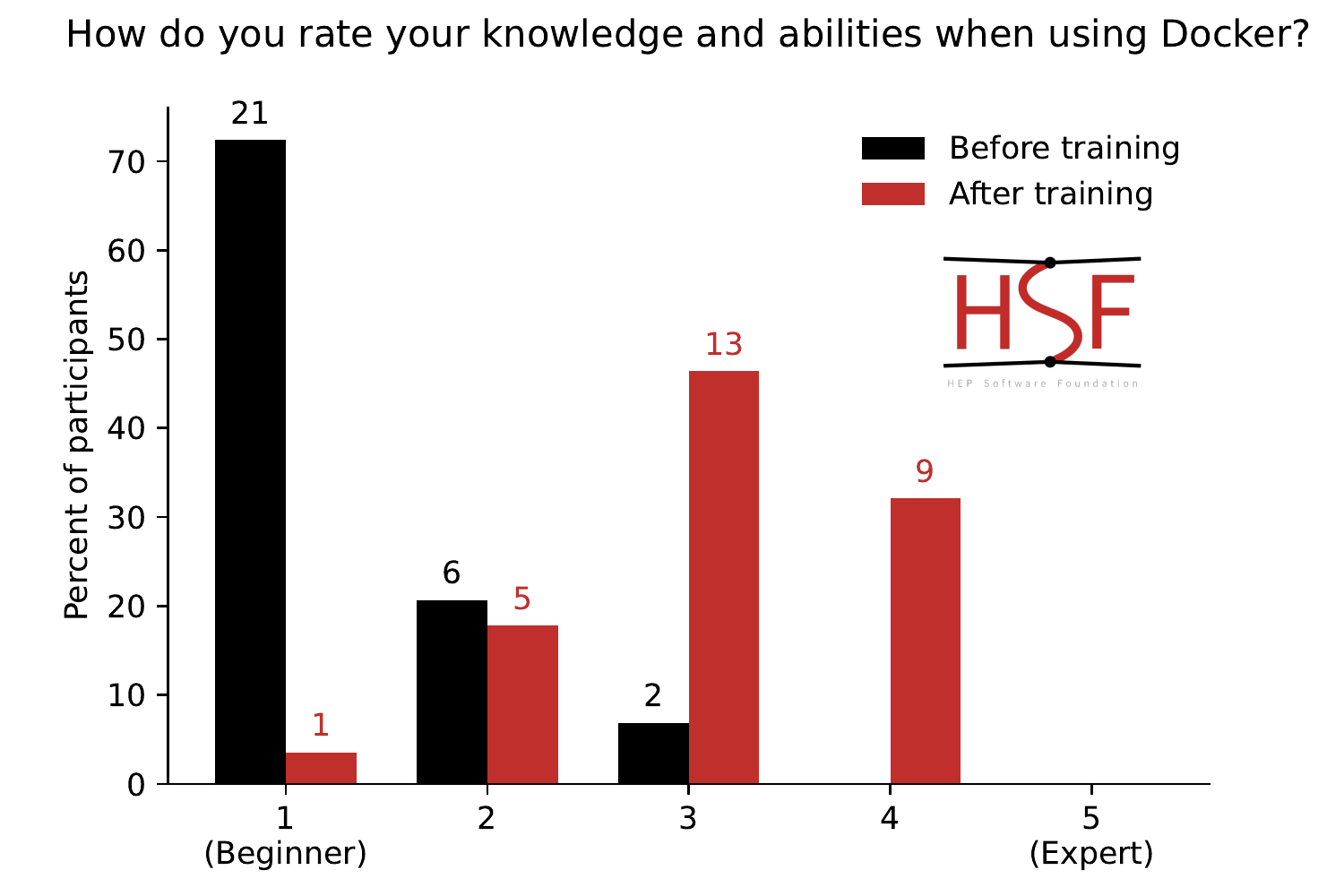}
\caption{The self-reported  pre- and post- training level of knowledge on the topic of Docker (a software container technology)
\label{fig:dockerfeedback}}
\end{figure*}

Figure \ref{fig:dockerfeedback} shows feedback on a training event involving containerization with Docker~\cite{HSF-training-Docker-2020}: clearly the training made a difference.  However, we are aware that this type of \enquote{learning evaluation} does not fully encompass the impact of our training. It only probes the perceived and self-reported learning of a skill. Instead, what is needed is a survey that is conducted sufficiently later to understand how well the learned skill is being applied in the context of research.
\section{Community}\label{sec:Community}
The solutions to future computing challenges require a large workforce trained in a wide range of software skills. In order to train this workforce, we rely on an active community whose members are enthusiastic and motivated to teach. Our members include people with various roles and backgrounds in HEP, such as experimental physicists from different collaborations, as well as software engineers from different institutes. As we scale our training activities, we also have members from nuclear physics and computer science as well. Members of The~Carpentries teach part of our very basic curriculum by an agreement via membership subscription through IRIS-HEP. The overall diversity of the background of the instructors and mentors adds great value to the training. Each educator brings their own flavor of experience from a different computing environment with a common goal of creating, teaching, and sustaining a common set of software skills.

As the success of our mission depends crucially on the motivation and participation of the community, we cultivate a strong sense of community ownership and pay special attention to acknowledge contributions of all kinds. We also encourage the participants in our training events to remain active or become more  active, share feedback, and in particular, to sign up to be a mentor in one of the next iterations of the same training module. If former participants do not yet feel confident about their mentoring skills, we offer to match them with a more senior mentor. In the same way, we encourage mentors to become instructors or facilitators and to become more and more active in our organization. By actively engaging participants and educators throughout the training community, we can sustain and nurture a culture of intentional learning and grow our community in an organic fashion~\cite{hsf-community-building}.

Educators not only provide an invaluable service to the HEP community, but they also get the opportunity to develop and sharpen their pedagogical skills and enhance their professional portfolio. About two-thirds of the HEP workforce eventually works outside of HEP, such as in the software industry and in data science. The training makes a meaningful difference in the preparation for such careers in terms of software knowledge and experience, and enhances the employability for both the educators and participants. The skills taught and learned, like Python, machine learning, and data analysis, align with the needs of the software industry and strengthen the job profile of a physicist to work in industry. At the same time, recognizing the importance of software skills within HEP may hopefully help to provide more incentives and clearer career paths in academia to those who want to pursue their career within HEP, or in other scientific research fields.
In particular, strengthening the research software engineer career path~\cite{rse_dan,rse_irishep} could significantly help retain the expertise within the HEP community.

\section{Sustainability}\label{sec:Sustainability}

Sustainable software~\cite{dan_katz} is essential for HEP.
A sustainable training program~\cite{sustainability_flashtalk} is key to pursuing this goal.
While continuing the existing work, it will be essential to  spread the training events and training expertise geographically to keep the costs low and move to an online training model to reduce financial burdens that accompany in-person training. In parallel, it is important that as the curriculum grows, it begins to include material specifically aimed at making software sustainable.

Training should be structured so that a minimal set of people are needed for maintenance and costs per event are minimized.
Growing the community is an important aspect of sustaining the workforce. Providing recognition and possible financial incentives can keep the community vibrant and motivated.
The community should recognize and appreciate the broader value of our software training, which prepares a workforce to solve computing challenges that are essential to advance our field and society at large.

To lead software training across HEP and related communities over the long run, we need a core team whose main focus is to support the overall mission of HEP software training. To scale up training efforts, we need to build mentorship and leadership at the local and regional level supported by the core team. Specifically, while we have started the following set of activities, we need to scale up by:
\begin{itemize}
\item Engaging more HEP labs, institutes, and universities in this endeavor.
\item Promoting equity, diversity, inclusion, and accessibility in participation across HEP communities and  being mindful of under-resourced institutions in different geographical regions.
\item Establishing a mechanism to get feedback from our communities and improve the training.
\item Ensuring that our core team and volunteers are afforded opportunities to grow professionally and have career paths.
\item Exploring ways to manage a financial support model to share costs in the long term.
\end{itemize}

\section{Broader Impacts}\label{sec:BroaderImpacts}

HSF-led training is multilayered, with a basic HEP software curriculum progressing to HEP-specific physics tools. Integrated with this is a growing outreach program that is essential to building an influx of software workforce and training young minds, catching them early in their educational development. For example, several outreach events are organised on introducing Python programming to K-12 teachers~\cite{HSF-training-DA-STEM-PR-2020} under IRIS-HEP and FIRST-HEP. The teachers can turn this into a classroom experience for their students where physics, astronomy, and math courses can have problem solving components that integrate programming with Python. In outreach events, the teachers analyze and interpret physics data with Python using Google~Colab~\cite{google-colab}, which allows them to work directly in the web browser without requiring any additional setup. Workshops teaching the basics of machine learning to school teachers are also organized~\cite{HSF-training-ML-STEM-PR-2021}. We plan to scale this experience by partnering with other stakeholders in HEP outreach, for example, Quarknet~\cite{quarknet}, which already has a well developed network of teachers and schools taking part in HEP outreach programs.

\section{Summary}\label{sec:Summary}

HSF and IRIS-HEP are creating software training and ensuring sustainability of software in HEP for years to come. The training material is open source and open access, shared publicly via GitHub. This allows anyone to join the discussion and make contributions by proposing changes, thereby continuously improving the available material.
This process is guided by continual feedback solicited from the participants of the training events. Finally, we have established a growing community of educators to broadly promote a culture within HEP that goes beyond valuing software skills, but also values the teaching of those skills to others. In doing so, we aim to foster a more active, inclusive, and diverse scientific community.  By leading software training across HEP and related communities, we will be able to meet the challenges in the field and beyond.

\section*{Declarations}
\begin{description}
\item[Funding:] This work is supported in part by National Science Foundation Cooperative Agreement OAC-1836650 and grants OAC-1829707 and OAC-1829729.
\item[Conflicts of interest:] The authors have no conflicts of interest to declare that are relevant to the content of this article.
\item[Availability of data and material:] Data sharing not applicable to this article as no datasets were generated or analysed during the current study.
\item[Code availabiliy:] Not applicable.
\end{description}
\bibliographystyle{spphys}       
\bibliography{references}   

%
%

\end{document}